# EFFICACY OF ATTACK DETECTION CAPABILITY OF IDPS BASED ON ITS DEPLOYMENT IN WIRED AND WIRELESS ENVIRONMENT

Shalvi Dave[1], Bhushan Trivedi[2] and Jimit Mahadevia[3]

[1]Department of MCA, Indus University, Ahmedabad
shalvidave.mca@iite.edu.in
[2]Director, MCA, GLSICT,Ahmedabad
bhtrivedi@yahoo.com
[3]Elitecore Technologies Pvt. Ltd., Ahmedabad
jimitm@yahoo.com

## ABSTRACT

*Intrusion Detection and/or Prevention Systems (IDPS) represent an important line of defence against a variety of attacks that can compromise the security and proper functioning of an enterprise information system. Along with the widespread evolution of new emerging services, the quantity and impact of attacks have continuously increased, attackers continuously find vulnerabilities at various levels, from the network itself to operating system and applications, exploit them to crack system and services. Network defence and network monitoring has become an essential component of computer security to predict and prevent attacks. Unlike traditional Intrusion Detection System (IDS), Intrusion Detection and Prevention System (IDPS) have additional features to secure computer networks.*

*In this paper, we present a detailed study of how deployment of an IDPS plays a key role in its performance and the ability to detect and prevent known as well as unknown attacks. We categorize IDPS based on deployment as Network-based, host-based, and Perimeter-based and Hybrid. A detailed comparison is shown in this paper and finally we justify our proposed solution, which deploys agents at host-level to give better performance in terms of reduced rate of false positives and accurate detection and prevention.*

## 1. INTRODUCTION

In order to apply admission and access control for a network, various Intrusion Detection and Prevention systems (IDPS) are available in the market. Intrusion detection system is used to manage traffic in real-time for increasing the accuracy detection and decreasing false alarm rate. In some instances, IPS adopts techniques from intrusion detection, such as detection approach, monitoring sensor, and alert mechanism. An IDPS is also used for gateway appliance, perimeter defence appliance, all-in-all capability, and network packet inspection/prevention. It is designed to identify and recognize potential security violations in stream network. However, the primary intrusion prevention use signature mechanism to identify activity in network traffic and host where perform detect on inbound – outbound packets and would block that activity before they access and damage network resources.

Fig.1 and Fig. 2 shows the basic scenario of an Intrusion Detection System (IDS) and an Intrusion Prevention System (IPS).

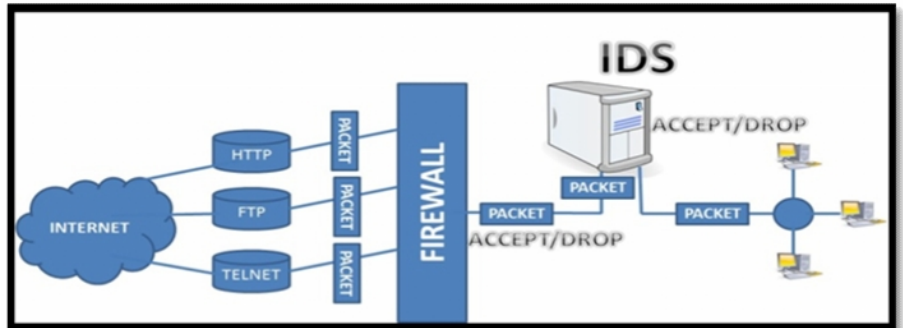

Fig.1    Intrusion Detection System

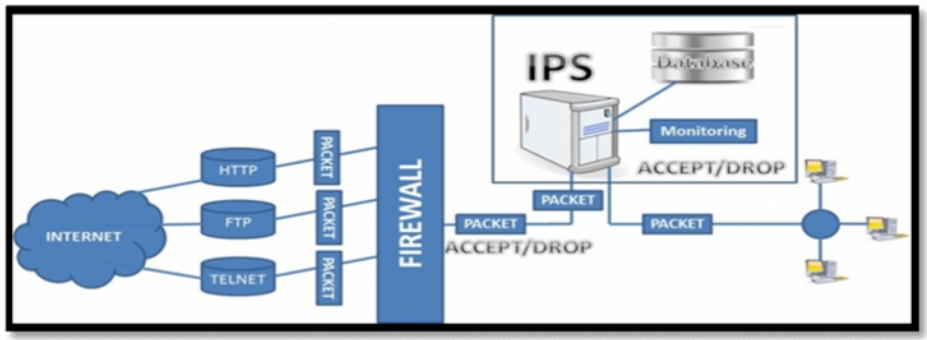

Fig.2    Intrusion Prevention System

An IDPS is an inline approach to monitor network activity. The detection technique used by the IDPS classifies it into two categories: signature based if it detects an attack by comparing it against a stored set of pre-defined signatures. It is anomaly-based if any abnormal behavior or intrusive activity occurs in the computer system, which deviates from system normal behavior. System normal behavior such as kernel information, system logs event, network packet information, software-running information; operating system information etc is stored into the database [1]

The deployment of an IDPS categorizes it as host-based or network-based. In addition, when deployed around the boundary of a network, it is known as perimeter-based IDPS. A distributed deployment of IDPS, wherein certain tasks are handled at the host-level and remaining at the network-level is known as hybrid IDPS. In this paper, we present a case study on the above mentioned techniques of deployment of existing IDPS, including problem areas faced in today's environment and enhancements possible to address each of these problem areas.

## 2. RELATED WORK

The primary deployment of IDPS is either at network level or at host level. The deployment determines the basic characteristics of an IDPS, which is then known as network-based IDPS (NIDPS) or host-based IDPS (HIDPS). In an NIDPS, an IPS sensor is usually placed at network ingress point. The IPS sensor monitors network traffic and inspect packet transmissions for suspicious behaviour. A network-based system can be used to provide detection for multiple hosts by locating the monitoring component appropriately (at a network ingress point, for example). HIDPS operate on single hosts, and operate on low-level system data, such as patterns of system calls, file access, or process usage. They can monitor for suspicious behaviour, or they can scan configurations to detect potential vulnerabilities using techniques such as port scan. Fig. 3 shows the deployment using HIDPS.

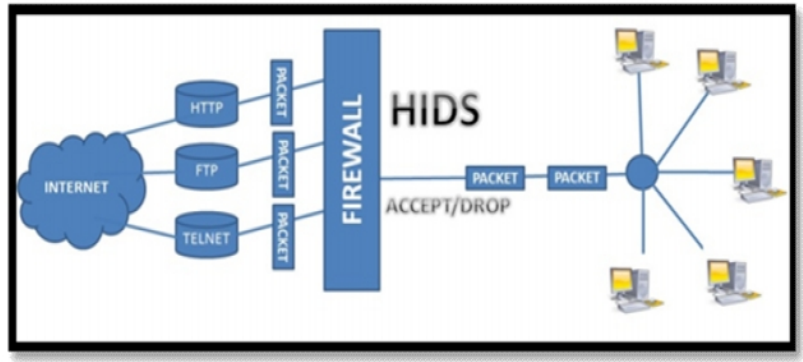

Fig.3 Host-based IDPS

Nowadays, the host-based approach plays a more prominent role than a decade ago. First, modern operating systems have grown in complexity, driven by the explosive growth of the Internet, thus it is more difficult to achieve an extensive monitoring. Secondly, system administrators are usually concerned about the impact of an HIDS on host performance. A notable HIDS (it is usually called a "web application firewall") is ModSecurity [9]. It is a module (i.e., a pluggable software component) for the Apache web server. ModSecurity intercepts incoming requests, runs the analysis and, in case a request is considered suspicious, can drop it, thereby preventing the request from being processed by the Apache instance.

The main advantage of the NIDS approach is the possibility to monitor data and events without affecting host performance. On the other hand, the fact it is not host-based turns out to be one of the main disadvantages (especially for systems analysing the payload of network packets). For instance, a NIDS cannot function properly in combination with applications or application protocols which apply data encryption (e.g. SSH and SSL), unless the encryption key is provided. A possible solution to this makes use of a host-based component to access data after decryption, but this causes an overhead on the monitored host. This problem is going to grow in importance since now IPv6 is gradually replacing IPv4: in fact, one of the main design goals of IPv6 is the authentication and confidentiality of data (through cryptography). Fig. 4 shows deployment using NIDS.

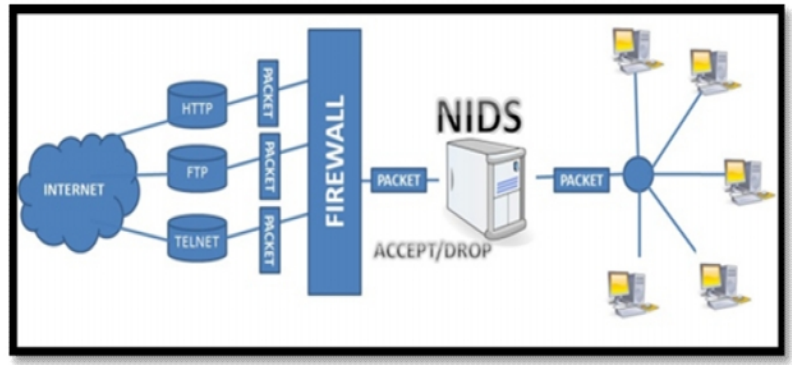

Fig.4 Network-based IDPS

Another common problem for a NIDS is the reconstruction of network traffic. Data streams are split into TCP segments and IP datagram. In order to analyse the content, the system needs to reassemble the traffic into the original form. Modern networks operate at high speed (up to 10Gbs): while the traffic reconstruction would be theoretically possible for an arbitrarily powerful system, a NIDS faces performance and implementation constraints. First, the NIDS must save a significant amount of data for a long time, depending on system time-outs and data throughput (this is resource consuming). Secondly, operating systems implement heterogeneous network stacks and handle data reconstruction differently. Therefore the NIDS engine should implement some context-awareness functionalities. All of these limitations resulted in the so-called evasion and insertion attacks, formalised by Ptacek and Newsham [10]. Attackers craft communications to fool the NIDS, e.g., by overwriting inside NIDS memory some data previously sent or by forcing the NIDS to drop data (that has not been analysed yet) after sometime.

The EMERALD system [3] attempted to merge the advantages offered by both the HIDS and the NIDS approaches into (virtually) single IDS. The problems of data normalization from different sources, event fusion and correlation and suitable metric definition are still open issues. These problems stopped the development of improvements after the first proof of concept of EMERALD.

There are further two types of NIDPS. Promiscuous-mode network intrusion detection is the standard technique that "sniffs" all the packets on a network segment to analyze the behavior. In Promiscuous-mode Intrusion detection & prevention systems, only one sensor is placed on each segment in the network. Network-node intrusion detection and prevention system sniffs the packets that are bound for a particular destination computer. Network-node systems are designed to work in a distributed environment [13].

In order to detect and prevent maximum number of attacks, we need to capture data that is distributed spatially and temporally. For example, an attack detected at different monitoring locations can be a distributed attack. Also, same attack that is detected during different time intervals gives an indication of co-ordinated, automatic attack. To capture data spatially and temporally, we require a hybrid system, which incorporates centralized monitoring feature of NIDPS and localized, distributed feature of HIDPS.

# 3. COMPARISON OF NIDPS AND HIDPS

This section describes deployment considerations to be taken while deploying an IDPS. We have studied few host-based and network-based deployment and show details of each with pros and cons. Host-based IDSs have certain advantages when compared with network-based intrusion detection systems. One advantage is that HIDSs can access semantically rich information about the operations performed on a host, whereas NIDSs that analyze network traffic have to reassemble, parse, and interpret the application-level traffic to identify application-level actions [10]. This is even more evident when application-level traffic is encrypted. In this case, a network-based monitor has to be equipped with the key material needed to decrypt the traffic; otherwise, the application-level information is not accessible. In addition, the amount of information that HIDSs have to process is usually more limited, because the rate at which events are generated by the OS and applications is smaller than the rate at which network packets are sent over busy links. A third advantage is that HIDSs are less prone to evasion attacks because it is more difficult to desynchronize the view that the intrusion detection system has of the status of a monitored application with respect to the application itself. Finally, a host-based intrusion detection system has a better chance of performing a focused response because the process performing an attack can sometimes be easily identified and terminated.

One of the more popular host-based intrusion detection and prevention systems is Hawkeye solution. [2] The architecture of Hawkeye solution proposed in [2] includes components such as sensors/agents, management server, database server, console and demilitarized zone (DMZ). This solution scores over other HIDPS by providing features such as capturing packets organized by TCP or UDP threads, passively monitoring network, packet viewing and logging in hex-format, detection of abnormal packet on comparison with benchmark ones and stating cause of abnormality. In case of abnormality, the source IP address can be traced. However, the basic detection methodology is packet-based. If an attack is distributed across multiple packets, it cannot be detected. Therefore, detection should be stream-based or data-based but not individual packet-based.

Another problem of Hawkeye solution is discussed [3][4]. It says that network flow identification should be done in such a way that every packet is monitored. To address this problem, an adaptive sampling algorithm is proposed. This algorithm predicts future behavior based on observed samples. It utilizes the weighted least squares predictor to select the next sampling interval. Inaccurate predictions by the weighted least squares predictor indicates a change in the network traffic behavior and requires a change in the sampling rate.

On the other side, this algorithm only looks for trends in network traffic and can detect few attacks such as DOS attacks. When we deploy detection system on host, an IDPS that monitors network activity can only measure trends in network traffic and thus detect attacks such as DOS.

A major trend found nowadays is co-existence of IPv4 and IPv6 networks. This is because the depletion of IPv4 addresses space. Due to the massive investment in IPv4, network and established lots of applications, and IPv6 networks need to be gradually perfect and recognized; the transition from IPv4 to IPv6 will be a very slow process [5]. The CIDP [5] is a proposed multi-level, distributed, three-dimensional architecture of intrusion detection and prevention. It consists of UTM (Unified Threat Management) for network-based intrusion detection and prevention systems UTM NIDP at the network boundary, the network-based intrusion detection and prevention systems Subnet NIDP in each subnet, Host-based HIDP, host-based mobile users HIDP in IPv6/IPv4 IPSec tunnel endpoint, host-based intrusion detection and prevention systems HIDP composition in the public domain server DMZ. The subnet NIDP incorporated in one of the three layers, detects back door attacks. Subnet NIDP protects other subnets from the host, which launches an attack, but the subnet itself is still vulnerable. For this protection, we require host-based IDPS. This proposed architecture [5] deploys HIDPS only on certain endpoints. Rather the deployment should be done on each host, since it protects the network not only from server-side attacks but also from client-side attacks.

Without NAT, the IPv4 address space would have been exhausted a long time ago. However, the translation of address/port by NAT affects different other applications since applications behind the NAT have no way to know what the real address/port used by the hosts [14,15].Whenever an attack is generated from within the network or a host is being victimized in the network, IP address of both attacker and victim is necessary. For this, IDPS must be re-examined to perform correctly with NAT. When an IDPS is deployed on a perimeter router, which is behind NAT device, we lack actual identity of attacker and victim. IDPS deployed in the network must be aware about the presence of a NAT device that changes the packets headers. [6] One of the solution to this [6] deploys two IDPS: one deployed above the NAT and another below the NAT, so these two systems will refer *attacker* and *victim* with different identities even if they alert the same attack, so these two alerts will be considered as two different alarms which increases the number of alerts and overwhelm the security operator. Identification module analyzes output of the analysis module to determine the real hosts' identities that are implicated in the security issue based on the NAT table.

To integrate IDPS to NAT box is tough and not every NAT device may provide such information outside the box. If we deploy IDPS functionality on a host, then the attacker and victim information will always be correct.

In case of wireless networks, because of some characteristics network, it is not so convenient to build an IDPS in wireless environment as in wired environment. For example, a company's trusted workers may need "inside" kinds of connectivity while using wireless devices. Inversely, visitors may need "outside" kinds of connectivity while connecting to the company's wired network through an access point inside the corporate firewall. It is very hard to place a firewall between "inside" and "outside". [16]

Secondly, the IDPS engine should be placed in sole path of user's traffics. But attacks on a wireless network can come from all directions and target at any node. Therefore, it is not easy to find a sole path to place an IPS engine that all traffic must pass.

Therefore, it is difficult to build an IPS engine in wireless environment as in wired networks. In order to address this issue, WBIPS (WTLS-Based IPS) model has been described [7]. In this mode, a logical sole path is built between every wireless terminal and its destination, so an IPS engine can detect and prevent the traffics of user.

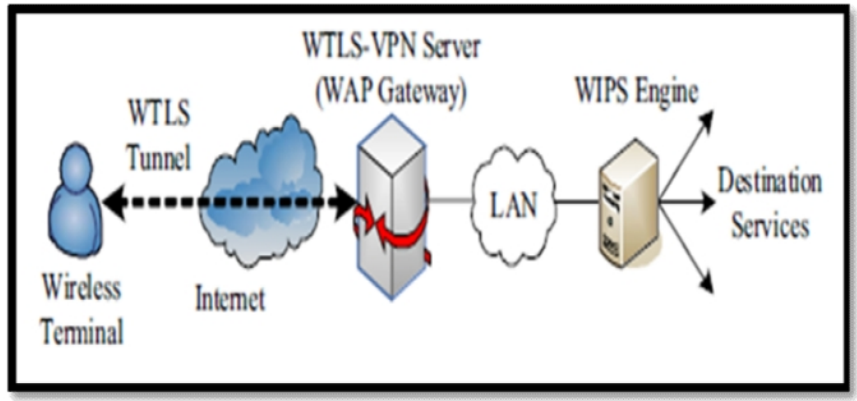

Fig. 5. Wireless Setup

VPN tunnel should be created between wireless device and gateway. The traffic should be re-directed through the tunnel. If we deploy HIDPS on wireless device, tunneling will not be necessary.

# 4. CONCLUSION AND THE PROPOSED SOLUTION

In order to detect and prevent maximum number of attacks, we need to capture data that is distributed spatially and temporally. For example, an attack detected at different monitoring locations can be a distributed attack. In addition, same attack that is detected during different time intervals gives an indication of co-ordinated, automatic attack. To capture data spatially and temporally, we require a hybrid system, which incorporates centralized monitoring feature of NIDPS and localized, distributed feature of HIDPS.

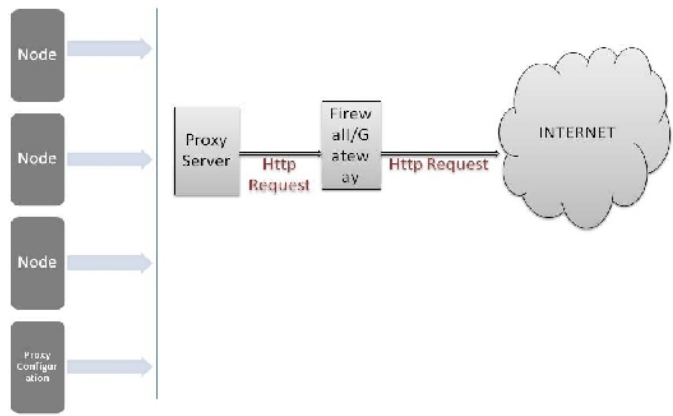

Fig.6 Typical Deployment of Proxy Server in Local Network

If someone wants to implement above-mentioned approach then it can better implemented using our work. As we have explained that implementing, it at edge router level might lead chances to run into a situation of false positives due to wrong operative frequency calculation. If the same can be implemented at host level then one can find out operating frequency of services easily using our Application Aware Logger System. Using our system one can get the hold on application data also to determine if the request is coming from proxy server as whenever any proxy server sends a request on behalf of any host we can find out actual host using "X-ForwardedFor" tag in request header. Using this parameter, we can implement operating frequency of concurrent connection in a better way.

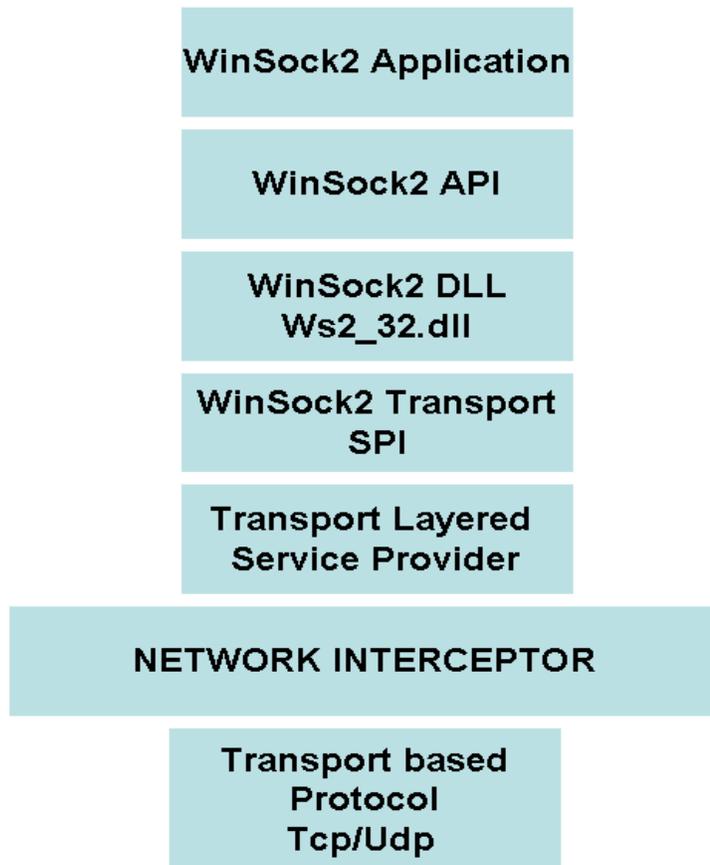

Figure 8. Implementation of Network Interceptor

In our proposed hybrid system, we have developed a module called Network Interceptor [16], which works to access real-time application data. It is shown in Figure 7. It captures socket calls using layered service provider of Winsock control and obtains all application related information and socket information. It monitors the data sent or received from each application using Suricata for attack detection and prevention. If any malicious activity has been found then it generates the event for the same, which includes not only source and destination information of connection but application information also like name of the application, version of the application, etc. For distributed monitoring, we have deployed a Logging agent in our hybrid system on each host. The logging agent captures events and sends the event information using UDP protocol, to Event Collector [17], which stores the log in database. Figure 8.shows how event collector works.

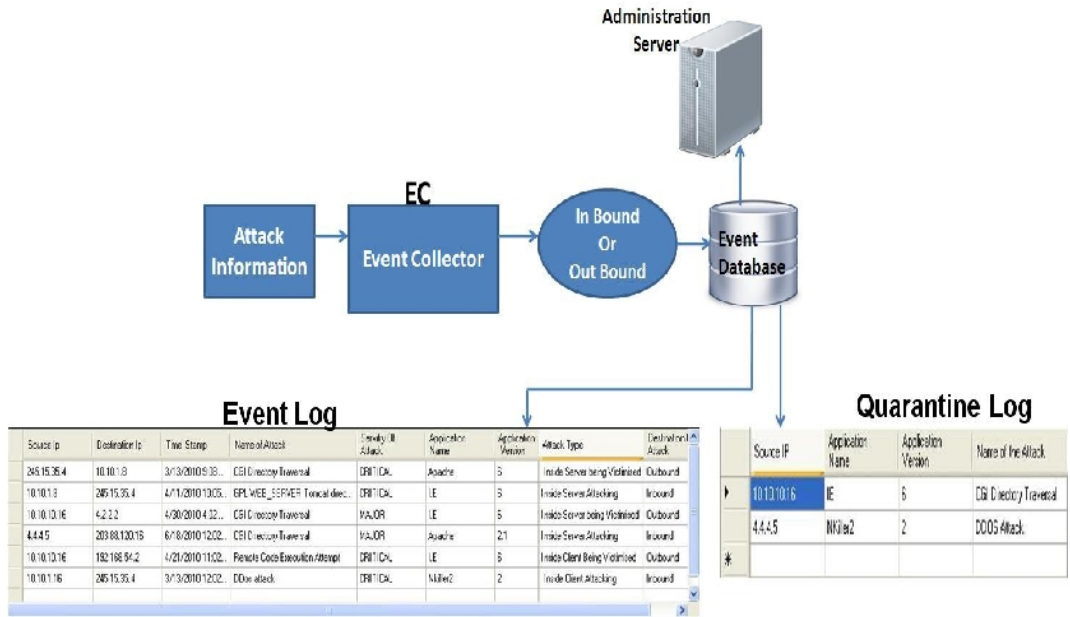

Figure 8: Role of Event Collector

First, we are using rule-set of Suricata, which is an IDPS used widely nowadays. From the existing rule-set of Suricata, we have taken two sets: Web-client and Web-server rules. Since our proposed system is designed taking into consideration corporate environment, we have classified the rule-set into further four categories:

1) Server-side Inbound.
2) Client-side Inbound.
3) Server-side Outbound.
4) Client-side Outbound.

This categorization is because an attack can be launched from within the network or from outside the network. In a typical network, there are two types of applications running: Client application and Server Application. Whenever a client application in the network requests for any service outside the network, it may become vulnerable to attacks from servers running outside the network. In addition, when any service provided by a Server application within the network is requested by an outside application, it may also launch an attack on server application. Apart from this, any vulnerable or infected application, client or server can possibly make attacks, to an application outside the network.

## 4.1 CLIENT-SIDE ATTACK

When any client accesses any service from server, there are high chances that an attack can be launched on client by the program running on remote server. This attack would generally be client specific. It is due to any known vulnerability of specific client version. We can categorize all such rules under Web-Client rules category and classify this type of attack as client-side attack. The client –side attack can be inbound as well as outbound depending on the direction of connection.

**A. Outbound Attack (Inside Client being victimized):** In this scenario, TOMCAT server is hosted on the internet. One of the desktop machines tries to access site hosted on the TOMCAT using Internet Explorer. If this site or server is being compromised then it can launch an attack on the internet explorer application. One such type of attack is "Possible Microsoft Internet Explorer URI Validation Remote Code Execution Attempt". If the version running of Internet Explorer is vulnerable to this attack then it is being victimized by such an attack. Our logging agent inspects the data and logs the event. The following diagram shows the actual working of the same:

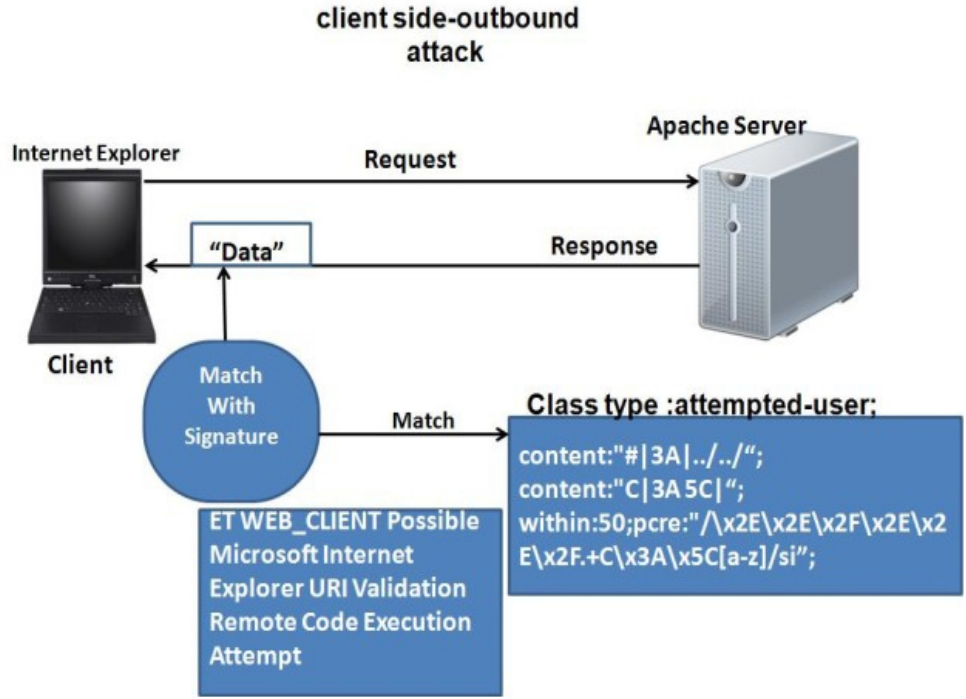

Fig 9: Client-Side Outbound Attack

**B. Inbound Attack (Inside server attacking on outside client):** In this scenario, TOMCAT server is running in our network. Some client from the internet tries to access the site using Internet Explorer. If our server is infected or compromised, then it can launch an attack on the remote client's IE. Our logging agent detects this attack and sends the log to central event collector. Event collector then stores this information into database. As it is the case of attack being generated within the network, so collector also adds this record into quarantine database. So in future, we can quarantine such infected applications running within the network.

## 4.2 SERVER-SIDE ATTACK

When anyone requests any service from the server, he can also land an attack along with the request. If running application server has any known vulnerability, services can suffer due to attack. We can have categorized all such rules under Web-Server rules category and classified this type of attack as server side attack. Again, depending on direction of connection, attack can be inbound as well as outbound.

**A. Outbound Attack**: A server is hosted on the internet. One of the desktop machines in our network tries to launch a DOS attack using an NKiller2 application. This application launches TCP Zero window attack on to remote server which is DDOS attack and due to that service gets interrupted. Our logging agent inspects this data, identifies an attack, and sends this data to event collector. Event collector stores this event log into database. Since it is an attack, which being generated from within the network Event collector also stores information into quarantine database.

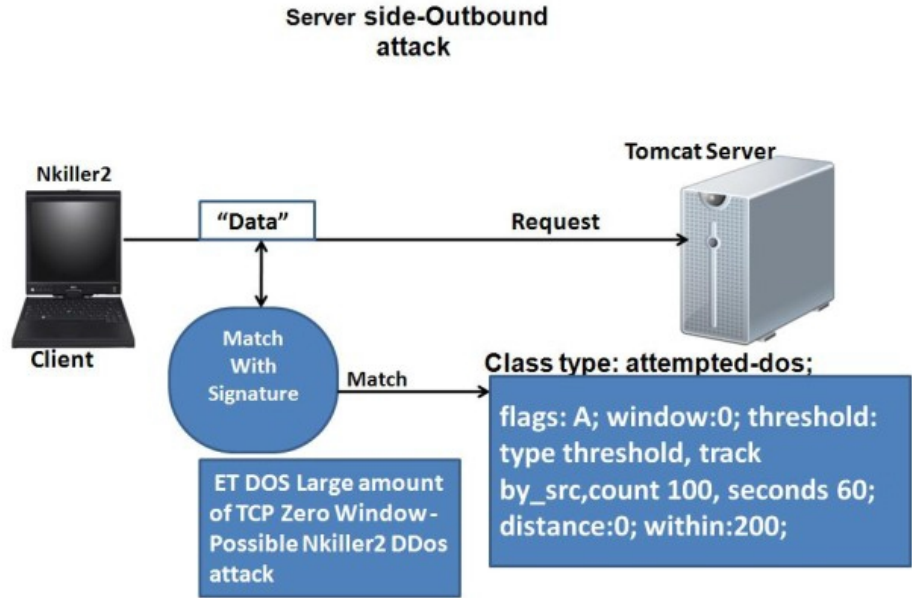

Fig.10: Server-Side Outbound Attack

**B. Inbound Attack:** In this case, TOMCAT server is running in our network. Client tries to attempt directory traversal on our server. Our logging agent logs event related information of such attacks and sends a log to central collector.

# 5. EVENT LOGGING FOR ADMIN USE

As described in previous section, logging agent with the help of network interceptor and Suricata inspects traffic to check whether exploit has been found. If found then agent sends an event to central event collector. This event log majorly identifies four different kinds of events. Inside client being victimized, Inside server being victimized, Inside client is attacking and Inside server is attacking. In case of attack generated by inside client or server collector also, add application related information like version, application name, and name of the attack into quarantine database. Administrator can also choose to apply any new security hot-fixes to application if available. As mentioned before we are using Suricata in our solution. Following is a brief description of how Suricata is used by our Event logger.

# 6. SURICATA

Our IDS logging agent inspects the data with the help of Suricata. Suricata is an open-source IDS available on all the platforms. It identifies an attack based on pre-defined signature rule-set.

Logging agent sends data along with application and connection information to Suricata. Therefore, Suricata does not need to track the connection for TCP Reassembly. After vulnerability scanning, IDS logging agent receives result back from Suricata. Result includes information about attack if detected along with its severity.

Suricata uses standard rule-set available from emerging threats. To achieve full security IDS signature rule-set has to be up to date. The logging agent contacts the administration server to check availability of new signature in the signature database. If new signature is found, logging agent also updates Suricata with new signature.

Conference on Advances in Computing and Artificial Intelligence, pg 29-33, ACM digital xplore, doi: 10.1145/2007052.2007059

## AUTHORS

**SHALVI DAVE** received the Master Computer Applications degree in 2001 from South Gujarat University, Surat, India. She is full time professor at department of MCA at Indus University, Ahmedabad, India. She is interested in Intrusion detection and Prevention Systems. E-mail: daveshalvi@yahoo.com

**Dr. BHUSHAN TRIVEDI** received his Ph.D in 2008 from North Gujarat University,India. He is working as Director of MCA department, GLSICT, Gujarat, India. His research interests include Intrusion Detection and Prevention Systems, Cryptography and Artificial Intelligence/ E-mail: bhtrivedi@yahoo.com

**MR. JIMIT MAHADEVIA** received his bachelor in computer engineering degree 1995. He is currently serving as Asst. Vice President, Elitecore Technologies Pvt. Ltd., Ahmedabad, India. His interests are Intrusion Detection and Prevention Systems, Wired and Wireless Network Security. Email: jimitm@yahoo.com